# On the thermal ("caloric") peculiarity of the SC transition, precursor to specific heat's "*jump*":

*its relation to the "paramagnetic" effect, precursor to the Meissner ejection – both revealed in HTSC material by the super-high sensitive SFCO method*

**Samvel G. Gevorgyan**[1,2] **and Bilor K. Kurghinyan**[1]

[1] Superconductivity & Scientific Instrumentation Center, Chair of Solid State Physics, Yerevan State University, 1 Alex Manoogian, Yerevan, 0025, Armenia
[2] Institute for Physical Research, National Academy of Sciences, Ashtarak-2, 0203, Armenia

**Abstract**
We have investigated the correlation between the "*paramagnetic*" peculiarity of the normal-to-superconductive phase transition – detected first in a *LTSC* tin, in 1989, and then confirmed in *HTSC* materials with much more higher resolution (*by a sensitive SFCO method*) – and the other weakly expressed effect, detected also both in *LTSC* & *HTSC* materials before their heat capacity's known "*jump*". In a low-$T_c$ superconductive tin this thermal effect is detected a half century ago, by *Corak*, but passed unnoticed so far. We show in this work that, externally similar these 2 fine effects really and truly have the same physical roots. To prove this assumption theoretically, and to reveal the reasons for their common origin, we use here a concept, which admits existence of two types of the Cooper pairs in *SC* materials – "*singlet*" & "*triplet*". As we conclude in this paper, they show completely different, non-traditional temperature behavior upon cooling of *SC* materials. We believe that this study results (along with the "*triplet*" superconductivity, proven experimentally in Josephson junctions quite recently, with the ferromagnetic barrier) may obtain key importance for the true understanding of the real nature of the superconductive phenomenon (*in whole*) and to uncover electron "pairing" mechanisms above the Meissner expel (*in particular*).

(Some figures in this article are in colour only in the electronic version)

---

## 1. Introduction

The problem of electron "*pairing*" above Meissner expel rose after 1986 only, when high-temperature superconductive (**HTSC**) materials were opened by *Bednorz* and *Muller* **[1]**. And, majority of scientists is admitted now **[2]**, that for truth there is need to consider two physical processes for *HTSC* materials – electron pairing, and onset of the phase coherence – separately, and independently of one another. So, superconductivity in *HTSC* material requires both the electron pairing & the Cooper-pair condensation. The later is also known as an onset of the long-range phase coherence among pairs **[2]**. In other words, it is admitted now **[2]**, that in *HTSC* material electrons become paired above Meissner ejection (above $T_0$, starting from $T_c$) and start forming the superconductive condensate only at the $T_0$, while in low-$T_c$ (*conventional*) *SC* material (**LTSC**), it is assumed that the pairing process and the onset of the phase coherence take place simultaneously (*at the same temperature*; $T_0 = T_c$): due to relatively large pair size, it is assumed that their wavefunctions are overlapping in a *LTSC* material.

Shortly after discovery of *HTSC* materials, however, a *"paramagnetic"* (**PM**) precursor to the superconductivity is detected by us in *LTSC* tin (**Sn**) **[3]** (**Fig.1**). Weakly expressed this effect (*in a central press it is reported later* **[4]**) & indirectly related to it another fine effect **[5]** (**Fig.2**) – a thermal (*"caloric"*) peculiarity, seen on heat capacity vs temperature curves, before the specific heat's known "*jump*" – provide together serious arguments, however, to hold a contrary opinion regarding low-$T_c$ *SC* materials. Though such a key effect (*also first detected in LTSC tin*) is available on *Corak*'s Figs. since middles of 50-s of the last century, however, to our surprise, it passed unnoticed for a long time. Seems, first we here focus attention to it. So, in this work we try to explain it, & list consequences following from it. Thus, we discuss below both the reasons why these 2 fine effects should have the same physical origin, and also, mark-out their key importance for true understanding of the real nature of the superconductive phenomenon.

To explain such a global similarity among thermal and electromagnetic properties of the *SC* materials (*regardless it is HTSC or LTSC*) we use here an advanced approach **[6-7]**, admitting existence of 2 types of Cooper pairs – both in *HTSC* and in *LTSC* materials – "*singlet*" and "*triplet*". Running ahead of the events let's note that, according to conclusions we come below, they show fully different non-traditional temperature behavior upon cooling of the material, starting with the birth of the first Cooper pair at temperature $T_c$. For *YBaCuO* composition *HTSC* material, respective curves (*plotted by formulas to be derived below*) are shown in **Fig.8**.

In regards to feasibility of separation of the electron pairing and Cooper-pair condensation, it was important searches for the said "*caloric*" effect in high-$T_c$ superconductive material. But, it is not so easy to measure heat capacity (*moreover, its little changes at beginnings of SC transition*) in so small volume clean *HTSC* objects (*such as film-structures, single crystals*). We did that recently, and could indirectly detect this fine effect also in a *YBaCuO* composition *HTSC* material **[8]** (see **Fig.3**) – by the step-by-step creation, perfection, and the proper use of a new principle of work imaging technique **[9-10]**, based on the combination of a well-focused laser scanning microscopy **[11]** and the **s**ingle-layer **f**lat-**c**oil-**o**scillator based highly sensitive new measurement method (**SFCO**–*technique*), introduced by our group for the high-resolution research in a last decade **[12-13]**.

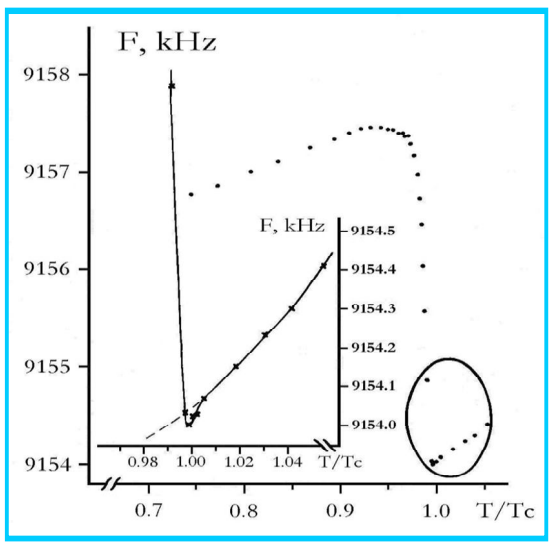

**Fig.1.** "*Paramagnetic*" effect detected upon transition to the *SC*-state of identical tin grains of ~ **5μm** in dia. **[3-4]**, *registered by the solenoid-coil based less sensitive technique*. **Inset:** enlarged view of the effect, broken line is device temperature dependence.

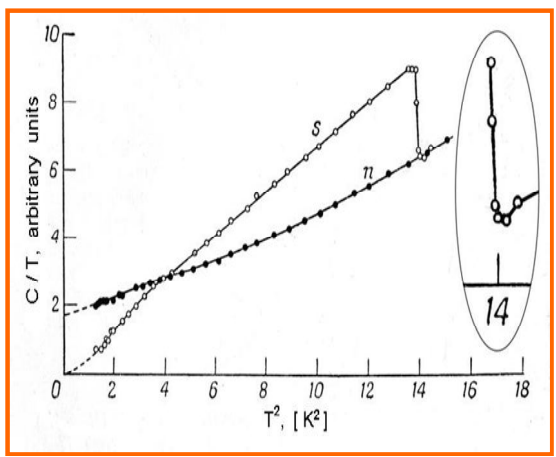

**Fig.2.** Heat capacity *vs* temperature curves detected in tin by *Corak* **[5]**. **Inset:** enlarged view of a fine effect seen before the specific-heat's "*jump*". Symbol "*s*" corresponds to the *SC*-state, while "*n*" – normal (*superconductivity is suppressed by magnetic field*).

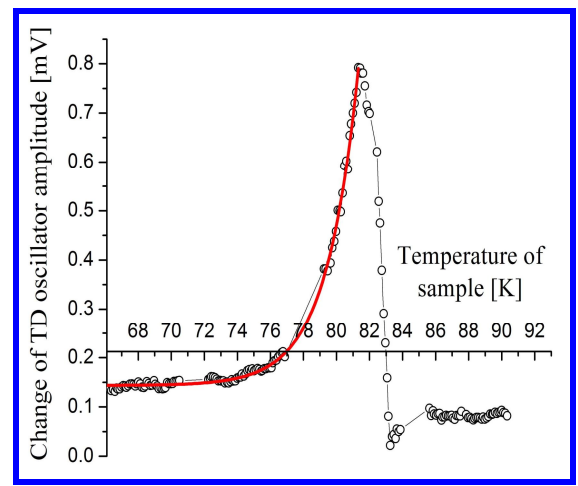

**Fig.3.** Indirect detection of the said "*caloric*" effect in *YBaCuO* **[8]** by a new imaging method **[9-10]**, based on the sensitive *SFCO* technique **[12-13]**. Circles are measured data, while the solid line – exponential fit of the exponential data (*compare this figure with the* Fig.2).

There are other *Meissner*-state precursor effects too. Before listing them let's remind, that superconductor is a double ideal material, since it becomes *ideal conductor* ($R=0$) and gets properties of the *ideal diamagnetic* ($B=0$) material below some temperature. The latter behaves also as an *ideal conductor* – reverse is not true. But, must the superconductor get such properties at the same moment (*temperature*)? And, why *different*-nature transitions (*1-st is assumed to be connected with electron pairing, followed by zeroing of the pair momentum,* *while the 2-nd, with pair condensation – due to collection of enough "singlet" pairs – proofs see in* [6]) should occur at a same temperature? And also, are there any other *different*-nature effects in Nature, which occur at a same moment? Such basic questions got the meaning since the said "*paramagnetic*" effect is detected in micron-size tin grains [3-4] ($T_c$~3.72K – **Fig.1**) that indicates the real physical onset of the Meissner expel. It precedes diamagnetic ejection, and substantially corrects shape of the normal-to-superconductive (*N/S*) phase transition curve. The origin of above questions relates also with a "*preceding*" effect, detected in percolating *YBaCuO* (in ceramics [1], and in films [14], with a granular structure of the material). According to it, *resistive*-transition ends before the start of Meissner expel. It was seen by *Morris* also in a *BiCaSrCuO* crystal [15], but doesn't attract a proper attention of the author – perhaps, due to lack of assurance in accordance among temperature scales of the conducted tests, performed in different setups. Questions were especially deepened when a "*diamagnetic activity*" was revealed in a *LaSrCuO* film by *Iguchi* [16] – at temperatures, much higher the transition temperature of the material, established by the onset point of the Meissner expel. A *super*-sensitive *scanning*-SQUID microscope is used for those tests. Such a flux activity is interpreted by the author as the effect, *precursor* to the Meissner state.

Listing *Meissner*-state precursor effects, let's stop also on the study conducted by *Valla* [17]. It shows that a "*pseudo-gap*" in the energy level of high-$T_c$ *SC* materials' electronic spectrum is the result of electrons being bound into Cooper pairs above the transition temperature to the *SC*-state, but unable to *super*-conduct, because pairs move incoherently. As to the *LTSC* materials (*which act much closer to the Zero*), it was said that, admittedly, superconductivity in these materials occur as soon as *electron*-pairs are formed. "In case of high-$T_c$ superconductors, however, electrons, though paired, "*do not 'see' each other* above some temperature," *Valla* says in [17], "and so, they can't establish the phase coherence, with all the pairs behaving as a '*collective*'". <u>Listed in this paper numerous data indicate, however, that such is the case for all types of the *SC* materials</u>.

And also, there is NO indication on any microscopic physical mechanism for establishment of the phase coherence among *SC* pairs in *Valla*'s works, as well as in papers of other researchers. While, as a hint to a possible mechanism for absence (*at higher temperatures*) & presence (*later, at cooling*) of the phase coherence among pairs might be discussed <u>existence of 2 types of the Cooper pairs in *SC* materials</u> (regardless it is high-$T_c$ or low-$T_c$) – "*singlet*" and "*triplet*", with different angular momentum [6]. As is indicating our study below, they have fully different temperature behavior upon cooling of the *SC* material (*qualitatively illustrated by* **Fig.8** *for the YBaCuO*). Such an approach may result in the *ideal conductivity* (it starts with the pair formation from the $T_c$, and acts for both types of the pairs). That is because the pairs are *quasi*-particles with a Zero momentum and so, with an infinitely large *de Broglie* wavelength, due to which they can move in a material without scattering, "*ignoring*" defects & impurities of the crystalline structure. At a later cooling of the material only, starting from the temperature $T_0$, such a reasonable approach may also result in the *ideal diamagnetism* (superconductivity, Meissner state) – but, for only "*singlet*" pairs…

And finally, a research team from the Ruhr-University (*Bochum*), Christian-Albrechts-University (*Kiel*) & *Santa Barbara* (headed by Profs. *Kurt Westerholt, Hartmut Zabel & Konstantin Efetov*) could make an experimental sensation recently: their studies on the "*pairing*" behavior of electrons have proven the "*triplet*" superconductivity [18]. In other words, they could detect existence of the pairs with parallel spin direction (for more details see [19]). The said integrated team has studied Josephson junctions with the barriers prepared from the Heusler compound $Cu_2MnAl$. In *as*-prepared state the $Cu_2MnAl$ layers are non-ferromagnetic, and the critical Josephson current density $j_c$ drops exponentially with thickness of Heusler layers $d_F$. On annealing the junctions at 240°C the Heusler layers develop ferromagnetic order and they could observe a dependence $j_c(d_F)$ with $j_c$ strongly enhanced, and weakly thickness dependent in the thickness range $7.0 < d_F < 10.6$ nm. The team interprets this feature as an indication of a "*triplet*" component in the *SC* pairing function, generated by the specific magnetization profile inside thin $Cu_2MnAl$ layers.

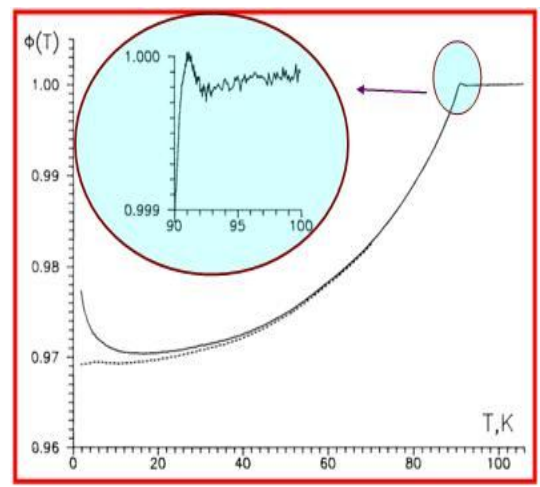

**Fig.4.** Superconductive phase transition of the ultra-fine *YBaCuO* powder with an average grain's size 2r=<u>**1920Å**</u> [20]. **Inset:** enlarged view of the "*paramagnetic*" (**PM**) effect.

Analysis of above data (*& some other results that we omit due to page limitation*) allow to expect, that sooner, there are no serious differences between *HTSC* & *LTSC* materials regarding the said two processes. Apparently, electron pairing & the Cooper-pair condensation (*phase coherence*) are separate & independent even in a *LTSC* material. Difference is in temperature scales. In *LTSC*, these processes run in a narrow range (*10-20mK* – **Fig. 1**), while in *HTSC* the scale is much longer. For example, for the *YBaCuO* composition *HTSC* material the scale of the "*paramagnetic*" effect (*that shows the scale of event selection for above 2 physical processes*), estimated by *Gantmakher*, is about *1K* [20] – **Fig.4**, while more sensitive our study, based on the use of *SFCO* technique, evidence that it is even broader *3K* [4] – **Fig.5**. Possibly, this is the reason why separation of the $T_c$ from $T_0$ was so problematic so far in *LTSC* materials. The problem is still open also due to lack of methods for the "*non-perturbing*", sensitive study of the *SC* transition in clean (*and so, tiny*) objects with small signals – especially at very beginnings of the transition, where even a highly sensitive *SQUID* technique is incapable to "*notice*" such small changes in a *normal*-state "*skin*"-depth (*comments on this matter see also in* [20]).

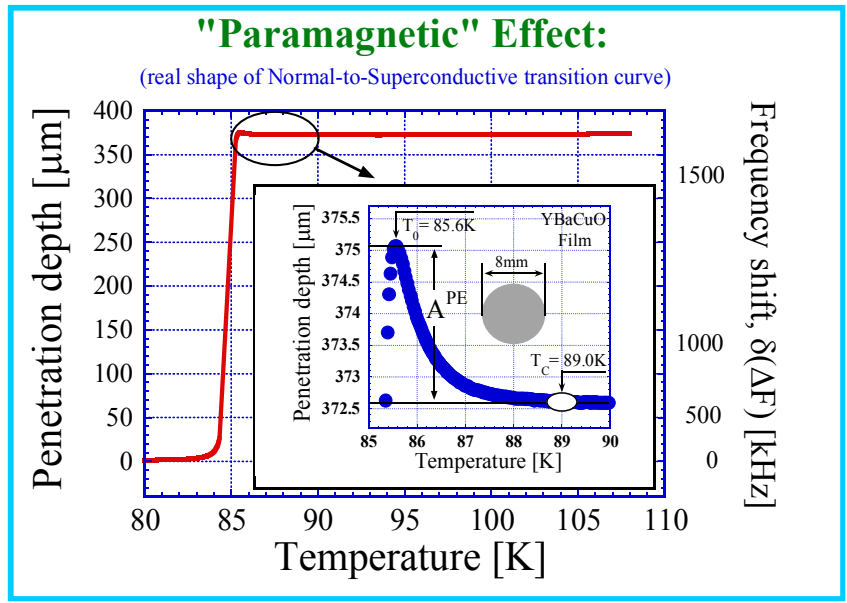

**Fig.5.** The normal-to-superconductive (**N/S**) phase transition of the *YBaCuO* composition *HTSC* disk-shaped film [4]. **Inset**: enlarged view of the "*paramagnetic*" effect (**PE**) at the very beginnings of the *SC*-phase transition, which precedes the *Meissner* expel. $A^{PE}$ is the effect's height.

## 2. Correlation between "*paramagnetic*" & "*caloric*" peculiarities of the superconductive transition

So, it has been shown by the measurements with a highly sensitive *SFCO* technique [12-13] that at cooling of *SC* materials there appears a new effect before the Meissner expel, named "*paramagnetic*" [4]. In other words, prior to the start of a diamagnetic expel at $T_0$, *SC* matter gets a fine paramagnetic nature, starting with $T_c$, and before the huge expel it slightly pulls-in pick-up coil's *MHz*-range testing radiofrequency (**RF**) field, resulting in a fall of a measuring tunnel diode (**TD**) oscillator frequency $F_{meas}$ (**Figs. 1** & **5**) – <u>that is why effect is named "*paramagnetic*"</u>. To be more precise note, that $\Delta F = F_{ref} - F_{meas}$ in **Fig.5**. Besides, frequency, $F_{ref}$, of the reference oscillator (used for compensation of a test device temperature dependence [12-13]) is independent of the effect, so, it remains constant during the tests. Note also, the effect is seen both in *LTSC* (**Fig.1**) & *HTSC* (**Fig.5**) materials – at cooling of the sample, and at its heating, as well. The effect is independent on the sample size or shape. But it gradually disappears in large objects, due to effect's averaging, due to temperature and/or material inhomogeneity in a sample volume – because the effect is seen in a narrow range (**Fig.1**). In this connection, let's state conditions when it was safely detected so far – that is important for true understanding of its nature. So, data related with the *PM* effect were obtained under conditions when the effect was independent on cooling/heating rates of a sample, and the *temperature*-stability & *material*-homogeneity throughout the specimen was high enough [4].

Along with this effect, combination of the *4-probe* method & our sensitive *SFCO* technique enabled to establish also disappearance of a resistance of *SC* materials before the start of the Meissner expel [6] – *resistive* transition ends ($R\cong 0$) before the start of Meissner expel. So, physical reasons for the *ideal conductivity* ($R=0$) & for the *ideal diamagnetism* ($B=0$) may not be the same in Nature. That is why, there are all grounds to believe, that there might be close relationships between the said "*paramagnetic*" & "*caloric*" precursors to the superconductive transition, and the *ideal conductivity* ($R=0$).

Below, we are going to give theoretical proofs why the "*paramagnetic*" effect [3-4, 20] (**Figs. 1** & **4-5**) detected prior to Meissner expel, and the "*caloric*" effect [5, 8] (**Figs. 2** & **3**) detected before the specific heat "*jump*" – both seen not only in *LTSC*, but also in *HTSC* – should have the same physical reasons & origin.

### 2.1 Theory

To explain a nature of the "*paramagnetic*" effect an idea admitting presence of 2 types of pairs is used by us in [6]. Analyses of the shape of effect & further development of this idea resulted in the creation of a phenomenological theory of the *SC* phase transition by *Sedrakian* [21]. In addition, in [6] it is concretized by us that, per-haps, some of pairs are "*singlet*" (with $\sigma=0$ spin), while remains are "*triplet*" ($\sigma=1$). That means, only later ones may contribute to the paramagnetism of the *SC* matter.

So, we suppose that in *SC* materials we actually deal with a *quasi*-particle system with 3 values of a spin – *I* = **0**, **1** & **1/2** (let's give this system a name **1-st model**). And, because of magnetic moment is an additive physical quantity, for such a 1-st model *SC* system we have:

$$M = (n_0/n)\cdot M_0 + (n_1/n)\cdot M_1 + (n_{1/2}/n)\cdot M_{1/2} \quad (1)$$

where, temperature dependences of the values for $n_{1/2} \equiv n_n(T)$, $n_0(T)$ and $n_1(T)$ are given by the formulas (2)–(4), (10) and (11) below, respectively.

If the Cooper-system wouldn't get a certain spin value at its formation (*that is, direction of electron spin does not depend on being of electron in a pair, or not*), then the spin values of the pairs of such a **2-nd model** system would not depend on temperature, and therefore, Cooper pairs would be the single-type at all temperatures. But, subject to both "*triplet*" a "*singlet*" pairs are available in a matter at starts of the *SC* transition (*with different behavior*), their amount should depend on temperature. Any other differences between these two kinds of Cooper pairs are not revealed empirically so far.

Now, our goal is to match these 2 model systems by comparing their heat capacities via their magnetic moments. For that it is enough to calculate expected difference of magnetic moments of the said model systems. By the energy difference of these 2 systems in the external magnetic field ( $E = -\Delta MH$ ) we will then calculate difference of heat capacities of these 2 model systems.

According to the above assumption and conclusion given by us in **[6]**, at beginnings of formation of the Cooper-pair system "*triplet*" pairs should be created first – that is because their stability is higher, so, their chance to be "survived" in a field of thermal fluctuations is higher. This certainly promotes entering of a testing *RF*-field into the sample – just which is registered by us in **Figs. 1** and **5** in the form of the "*paramagnetic*" effect. Let's also note that "*singlet*" pairs (*without a spin*) in no way may contribute to the paramagnetism of *SC* matter. While, experimental data & reasoning, presented in **[6]** & **[21]** lead to the suggestion, that further formation of the "*singlet*" Cooper-pair system in a coherent state, seems, is responsible for the establishment of the *ideal diamagnetic state* in a *SC* matter (*in other words, only the "singlet" pairs may result in Meissner repulsion*).

So, according to our conclusion in **[6]** (*based on the shape of a measured transition curve* – **Fig.5**) one may say, that due to formation of a Cooper-pair system concentration of "normal" charge carriers (*electrons*) in a matter decreases (*starting already from the 1-st critical temperature*, $T_c \sim 89K$), and at approach to the *2-nd* critical temperature, $T_0 \sim 85.6K$, it already drops to almost the zero (*to be more precise, to $n_{res}(T_0)$ – see below*) – in contrast to the traditional "two-fluid" models of the superconductivity offered in due time first by *Gorter* &

*Casimir* **[22]**. While, following by shape of the fine "*paramagnetic*" effect (**Fig.5**) one may conclude that the Cooper-pair system (*formed due to step-by-step reduced electrons*) not only is still unable to eject detecting *RF*-field of the pick-up coil from sample (*which is too weak in our method, & "non-perturbing"* **[12-13]**), but, it still permits the field to continue slowly "*intrude*" the sample. This process reaches to its peak when the specimen cools down to $T_0 \sim 85.6K$. And only below this point temperature dependences of the "normal" charge carriers (*electrons* – $n_n(T)$) and Cooper pairs, $n_s(T)$, approximately follow the rules of conventional theories **[22-23]**.

Temperature dependence of "normal" charge carriers, $n_n(T)$, near the *fluctuition*-temperature region may be presented by the following empirical formulas **[6]**:

$$n_n(T) \cong n_{res}(T_0)(T/T_0)^2, \quad \text{at } T < T_0, \quad (2)$$
$$n_n(T) \cong n\,[(T - T_0)/(T_c - T_0)]^2 + n_{res}(T), \text{ at } T_0 < T < T_c, (3)$$
$$n(T) \cong n, \quad \text{at } T_c < T, \quad (4)$$

where $n$ is the total carrier density.

To get an analytic function for further calculations let's seam together formulas (2)–(4). In agreement with requirements of continuity and smoothness of the function $n_n(T)$, we suggest for the residual density of "normal" charge carriers $n_{res}(T)$ a linear expression $\beta(T_c - T)$. Note, that $n_{res}(T)$ is negligible in *LTSC* material. But, as follows from variety of works **[24-27]**, in *HTSC* materials, being small in general it becomes noticeable in many important practical applications of these materials. So, for $\beta$ one may take, for example, the value $\beta = 0.01$ (*it may be shown, that $\beta$ can't be more than 0.06, in general* **[28]**). In that case, the function graph for normalized electron concentration, $\eta_n(T) \equiv n_n(T)/n$, in a temperature range $T_0 < T < T_c$, is shown in the **Fig.6**.

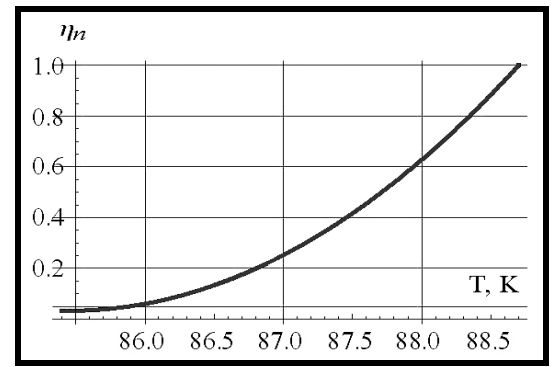

**Fig.6.** The graph for the normalized electron concentration, $\eta_n(T) \equiv n_n(T)/n \cong [(T - T_0)/(T_c - T_0)]^2 + n_{res}(T)/n$, in a temperature range $T_0 < T < T_c$ – for the case, when $\beta = 0.01$.

After seaming Eqs. (2)–(4), for $\eta_n(T)$ we finally get following formulas – for the whole temperature range

$$\eta_n(T) = \frac{n_{res}(T_0)}{n}\left(\frac{T}{T_0}\right)^2, \qquad \text{at } T < T_0, \quad (5)$$

$$\eta_n(T) = \left(\frac{T-T_0}{T_c-T_0}\right)^2 + \frac{n_{res}(T)}{n}, \quad \text{at } T_0<T<T_c, \quad (6)$$

$$\eta_n(T) = 1, \quad \text{at } T_c<T, \quad (7)$$

Hereafter we will deal only with these formulas. The full seamed curve for normalized electron concentration is shown in **Fig.7a**. The function describing concentration of all paired electrons (*Cooper pairs* – **Fig.7b**) is

$$\eta_S(T) \equiv n_S(T)/n = 1 - \eta_n(T) \quad (8)$$

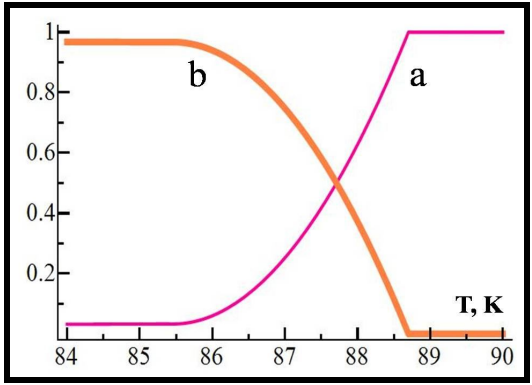

**Fig.7.** Graphs of the functions $\eta_n(T) \equiv n_n(T)/n$ (**a** – normal electrons), and $\eta_S(T) \equiv n_S(T)/n = 1-\eta_n(T)$ (**b** – Cooper pairs).

As was mentioned, based on *high*-resolution empirical data on the "*paramagnetic*" effect **[4]** (inset curve in **Fig.5**), a phenomenological theory of *SC* transition is created in **[21]**. Starting with an idea of the existence of 2 types of pairs in *SC* materials, it permits to calculate temperature dependence of a portion of "*singlet*" pairs, $\alpha(T)$, with respect to the total number of Cooper pairs.

$$\alpha(T) = \{1 - th[3.5 \cdot (T-83.5)]\}/2. \quad (9)$$

Thus, for concentrations (*relative amounts*) of "*singlet*" & "*triplet*" Cooper pairs in *SC* material we obtain

$$n_0(T)/n = \eta_S(T) \cdot \alpha(T)/2, \quad (10)$$
$$n_1(T)/n = \eta_S(T) \cdot (1-\alpha(T))/2, \quad (11)$$

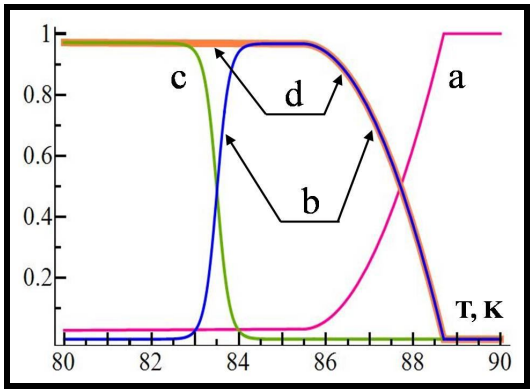

**Fig.8.** Temperature dependances of "normal" electrons (**a**), the "*triplet*" (**b**) & "*singlet*" (**c**) pairs, and the all Cooper pairs (**d**).

Symbols "**0**" & "**1**" are used to indice the spin of pair. By the use of formulas (5)–(7), (10) and (11), temperature dependences of normalized "normal" electrons (**a**), "*triplet*" (**b**) & "*singlet*" (**c**) pairs, and also all Cooper pairs (**d**) are plotted in **Fig.8**. As is seen from the figure, at cooling, starting with the beginnings of the *N/S* transition ($T \sim T_c=89K$) amount of "*triplet*" pairs (curve **b**) is dominant – and up to the $T \sim 83.5K$. And their amount reaches to its maximum just around the peak of "*paramagnetic*" effect ($T \sim T_0=85.6K$ – see inset in **Fig.5**).

Now it is time (*we are ready*) to discuss contribution to the magnetic moment (*magnetization*) & the heat capacity of the *SC* material, conditioned only by spins of "normal" electrons and Cooper pairs, and calculate their temperature dependences. As the real model for the further calculation let's take above defined 1-st model (*material consists of 3 quasi-particles with a following values of the spin – I = **1/2**, **0**, & **1***), for which amount of each type of particles depends on temperature according to above formulas (2)–(4), (10) and (11) – respectively for the "normal" electrons, "*singlet*" and "*triplet*" Cooper pairs. Besides, to have the same temperature everywhare inside the sample, the cooling and heating of the sample was performed too much slowly during the measurements (see the **Fig.5**, for example, otherwise the "*paramagnetic*" effect may dissapear, due to averaging). That permits to consider the problem under study almost quasistatic, and use advantages of the related calculation methods – to simplify further calculations.

To determine the magnetization of an ideal gas in the external magnetic field, consisting of particles with the given *I* value of the spin, let's suppose, that particles of such a model system are also identical and non-interacting. The statistic sum, $Z$, of such a system is

$$Z = \left(\frac{sh\left[\left(I+\frac{1}{2}\right) \cdot \frac{\mu_B \cdot g \cdot H}{kT}\right]}{sh\left[\frac{1}{2} \cdot \frac{\mu_B \cdot g \cdot H}{kT}\right]}\right)^n,$$

and, magnetization of a unite volume of such a model gas system in an external magnetic field, $H$, is **[28]**

$$M_I = n \cdot I \cdot \mu_B \cdot g \cdot B_I\left(\frac{I \cdot \mu_B \cdot g \cdot H}{kT}\right), \quad (12)$$

where, $B_I$ is the *Brillouin* function,

$$B_I(x) = \frac{2 \cdot I + 1}{2 \cdot I} cth\left(\frac{2 \cdot I + 1}{2 \cdot I}x\right) - \frac{1}{2 \cdot I} cth\left(\frac{1}{2 \cdot I}x\right).$$

In this formulas $g$ is the *Lande* factor, $\mu_B$ – the *Bohr* magneton, $H$ – the applied external field, $n$ – total number of electrons, and the $k$ is the *Boltzmann* constant.

Thus, we have a system of *quasi*-particles with 3 different values of the spin – $I = **1/2**$, **0**, & **1**. Besides, magnetic moment is an additive physical quantity, so magnetization of such a 1-st model *SC* system is given by the formula (1) above, where, as is mentioned, temperature dependences of the values $n_{1/2}$, $n_0$ & $n_1$ are given by the formulas (2)–(4), (10) & (11), respectively.

So, after necessary calculations **[28]**, for magnetization of the 1-st model system we obtain the next formula

$$M_{1-st} = n \cdot \mu_B \left\{ \eta_n(T) \cdot th \frac{\mu_B H}{kT} + \frac{\eta_s(T) \cdot [1-\alpha(T)]}{2} \times \right.$$
$$\left. \times g \cdot \left( cth \frac{3\mu_B gH}{2kT} - \frac{1}{2} cth \frac{\mu_B gH}{2kT} \right) \right\} \quad (13)$$

Similarly, for the magnetization of the 2-nd model system one may easily obtain the following formula [28]

$$M_{2-nd} = n \cdot \mu_B \cdot th \frac{\mu_B H}{kT}. \quad (14)$$

We remind, that our final goal is to match these two model systems by matching their heat capacities via their magnetic moments. So, expected difference of moments (*magnetizations*) of these 2 model systems is

$$\Delta M = n \cdot \mu_B \left\{ \eta_S(T) th \frac{\mu_B H}{kT} - \frac{\eta_S(T) \cdot [1-\alpha(T)]}{2} \times \right.$$
$$\left. \times g \cdot \left( cth \frac{3\mu_B gH}{2kT} - \frac{1}{2} cth \frac{\mu_B gH}{2kT} \right) \right\} \quad (15)$$

**Fig.9** demonstrates temperature dipendence of the function (**15**). Identity of the curve (**b**) in **Fig.8** with the curve in **Fig.9**, & their qualitative coincidence with the measured data on "paramagnetic" effect (*see inset curve in Fig.5*) witness that, apparently, microscopic reasons generating the fine "*paramagnetic*" effect retate to dominant role of the "*triplet*" pairs over the "*singlets*" at very beginnings of the formation of the superconductive state. Especially as because the peak of the "*paramagnetic*" effect is just at the temperature ($T_0$=85.6K), at which concentration of the "*triplet*" pairs reaches to its maximum possible value (*compare the* **Figs. 5** and **8**).

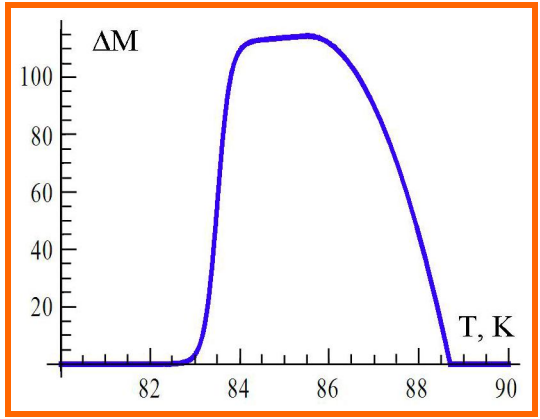

**Fig.9.** Temperature dependence of the diference of magnetic moments (*magnetizations*) between the 2 discussed model systems, *ΔM*, conditioned by the spins of the electrons, and the Cooper pairs.

The energy difference of these 2 model systems in the external magnetic field $H$ is given by the following formula, $E = -\Delta M H$, which allows to determine the difference of the heat capacities of these 2 model superconductive systems by the expression below

$$\Delta C_H = -H \left( \frac{\partial (\Delta M)}{\partial T} \right)_H. \quad (16)$$

After simple, but long calculations one may get for the function $\Delta C_H(T)$ the following analytic formula

$$\Delta C_H = \frac{1}{2} \mu_B \left\{ -\frac{2\mu_B H}{kT^2} \cdot \frac{\eta_S(T)}{[sh(\mu_B H/kT)]^2} - \right.$$
$$-\frac{\mu_B g^2 H \left( \left( \cos ech \left[ \frac{\mu_B H}{2kT} \right] \right)^2 - 6 \cdot \left( \cos ech \left[ \frac{3\mu_B gH}{2kT} \right] \right)^2 \right)}{4kT^2} \times$$
$$[\alpha(T)-1] \cdot \eta_S(T) +$$
$$+ g \cdot \left( \left( -\frac{1}{2} cth \left( \frac{\mu_B gH}{2kT} \right) + cth \left( \frac{3\mu_B gH}{2kT} \right) \right) \cdot \eta_S(T) \cdot \frac{\partial \alpha(T)}{\partial T} + \right.$$
$$\left. + 2th \left( \frac{\mu_B H}{kT} \right) \cdot \frac{\partial \eta_S(T)}{\partial T} \right) -$$
$$\left. -\frac{g^2}{2} \cdot \left[ cth \left( \frac{\mu_B gH}{2kT} \right) - 2cth \left( \frac{3\mu_B gH}{2kT} \right) \right] \cdot [\alpha(T)-1] \cdot \frac{\partial \eta_S(T)}{\partial T} \right\}$$
$$(17)$$

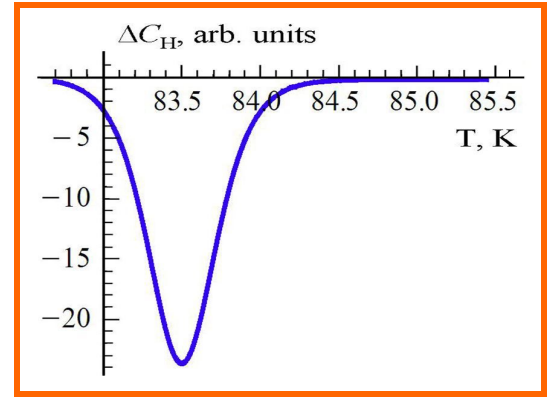

**Fig.10.** The graph of the function $\Delta C_H(T)$.

### 3. Discussion

So, we have compared magnetic moments (*magnetizations*) of 2 said model superconductive systems conditioned by only their spins. Then, we calculated difference of their heat capacities, $\Delta C_H$, conditioned by different magnetizations. As it turned out, that is enough for description of the fine "*caloric*" effect, shown in **Figs. 2-3**, detected both in *LTSC* & in *HTSC* materials, before their heat capacity "*jump*". **Figure 10** shows temperature dependence of the function $\Delta C_H(T)$, plotted by the formula (**17**). Qualitatively, it explains the fine thermal effect (*seen prior to the heat* capacity "*jump*"), detected in *LTSC* tin by *Corak* 50 years ago [5] (**Fig.2**). It is ignored in due time: perhaps, such a key effect is considered as a noise-level signal. Theoretical curve in **Fig. 10** agrees also with a similar "*caloric*" effect detected by us in *YBaCuO* by a sensitive *SFCO* method [8] (**Fig.3**).

Note that a qualitative explanation of the said thermal effect ("*paramagnetic*" effect as well) stand possible only after we have classified (*sorted*) the Cooper-pair system by the spin at our calculations: the "*singlets*" and "*triplets*", amount of which sharply changes in so complicated fluctuation temperature region at cooling.

Results obtained here allow to conclude, that a weakly expressed "*paramagnetic*" effect, detected recently & investigated in details by the created highly sensitive *SFCO* new test-method, and the no less fine peculiarity before the specific heat's "*jump*", detected long ago by *Corak & Satterthwaite*, have the same physical reasons – both of the effects result from the different spin values of the Cooper pairs and their different temperature behavior upon cooling of the superconductive material.

### 4. Conclusion, final remarks, and suggestions

Analyzing all discussed above, one may conclude that at the moment becomes important unambiguous demonstration of the "*singlet*"-to-"*triplet*" conversion of Cooper pairs (*and back*), and separation of the *ideal conducting* ($R=0$) and *ideal diamagnetic* (*superconductive* – $B=0$) states from each other. That is too much important not only for the true interpretation of the real nature of the superconductive phenomenon (*in whole*), but also, for the correct understanding of the microscopic mechanisms of the electron "*pairing*" (*the physical mechanisms for the establishment of the long-range phase coherence among the superconductive pairs*).

In this connection, one of the most effective experimental ways is investigation of the ***F**ulde-**F**errell-**L**arkin-**O**vchinnikov* superconductivity (the **FFLO** state **[29-30]**) – as sensitive, as it is possible (*and, many-sided*). One of materials, in which theorists suggest to study this unique phenomenon, is the heavy-fermion superconductor $CeCoIn_5$. First, that was experimentally investigated by *Radovan* et al. **[31]**, but, with not enough resolution, compared to what we suggest to do below. This material satisfies a delicate balance of properties needed to detect and study the FFLO superconductivity at temperatures below the 350mK. A highly sensitive *SFCO* test-method **[12-13]** (*with its unprecedented capabilities*) one needs to use as a unique scientific research instrument for precision study of the peculiarities of the superconductivity, and the pair formation in FFLO state. Such a research (*to be carried out approximately by the scenario implemented in a pioneering experiment* **[31]**) may enable direct demonstration of the "*singlet*"-to-"*triplet*" spin-flip of the Cooper pairs, and also, the separation of the *ideal conducting* ($R=0$) *and ideal diamagnetic* ($B=0$) states.

At this, distortion of the testing *RF*-field configuration near the flat coil face (*taken by the frequency rise of the testing SFCO technique*) one may use for the checking whether the *ideal diamagnetic* state (with $B=0$) is established or not. While, amount of the absorption of the testing *TD*-oscillator's power by the sample (*taken by the amplitude fall of the measuring SFCO technique*) may be used to check whether the *ideal conducting* state (with $R=0$) is established in a material, or not. Along with this, establishment of an *ideal conducting* state may be checked independently by the *4-probe* test technique.

Besides, our research shows that a flat-coil based oscillator can be activated also with its internal capacitance **[32]** (*without an external capacitance in its resonant circuit*). That is the result of a relatively high value of the internal capacitance of single-layer flat coils, compared to their parasitic capacitance with respect to the surrounding radio-technical environment. This key circumstance opens exotic areas for the flat-coil oscillator application. Namely, a "*needle-like*" measuring *MHZ*-range magnetic field of such a flat coil **[33]** (*used as a pick-up in such a stable TD-oscillator*), enables a novel method (*new approach*) for the surface probing **[34]**. Such an unusual probe shows strong dependence of a detected signal both on the lateral position of such a probe with respect to the surface of the object. and on the size of a spatial-gap between the probe and the surface of the object **[34]** – crucial for the probe microscopy. This opens an opportunity for the designing of the non-solid-state "*magnetic-field*" probes with the *RF* power applied to the sample, lying in the power range of 1nW to 5μW. The gap between such a "*probe-formative*" flat coil and the object can be larger than 1mm **[33-34]**, compared with ~1nm gap of acting tunneling and atomic force probe-microscopes **[35-38]**.

That is why, we believe that such a <u>*SFCO*-probe</u> may enable to distinguish (*both by the amplitude and by frequency of the detecting TD-oscillator*) details of the relief of the FFLO "*node*" structure, in a real space (*consisting of the alternating regions of the superconducting layers and the spin-polarized magnetic walls*). However, for that aim, we suggest to create and use a *SFCO* method-based advanced "*magnetic-field*" probe, with a lithographically made single-layer flat coil of about 1mm in diameter **[39]** – *as an effective "needle-type" probing instrument with better than* 100nm *predicted lateral resolution*. Such a new probe will have considerably large work-distances (*approx., 0.1-1 mm*) between the probe & the surface of the object, which enables a "*laser*" control of the local probing area of the object, and, if needed, application of the test and control perturbations.

### Acknowledgments

This study is supported by the Armenian **NFSAT** (*National Foundation of Science & Advanced Technologies*) and the US **CRDF** (*Civilian Research & Development Foundation*) under the <u>***Grants # ISIPA 01-04 and # UCEP 07/07***</u>. The study was partially supported also by state sources of the Republic of Armenia in frames of the task program on "<u>*New Materials*</u>" and *R&D* projects ## 01-431, 1359, 301-0046 & 72-103.

Besides, SGG is grateful to Prof. *V.F. Gantmakher*, for his comments on "*paramagnetic*" effect, opened first in tin grains 2 decades ago, which stimulated further studies of the effect in *HTSC* films discussed. He is grateful also to Profs. *M. Takeo*, *K. Funaki*, *T. Matsushita* & Dr. *T. Kiss* for the given chance to work

& conduct research in Kyushu University, which promoted & accelerated assembling of experimental data on this problem (*jointly published earlier*), used here for discussion & motivetion of the new ideas. SGG appreciates also discussions of test-data and the problem as a whole with Acad. *D.M. Sedrakian*, Profs. *E.G. Sharoyan*, *A.A. Polyanskii* & Dr. *N.M. Dobrovol'skii*. He thanks also to *V. Gevorgyan*, *H. Shirinyan*, *G. Karapetyan* and *S. Muradyan* for the all-round help during the preparation and at conducting of experiments, and for assistance at the paper preparation.

The authors are also grateful to Dr. *V.R. Ohanyan* for his help in optimization of the methods of calculation.